\documentclass[aps,prl,superscriptaddress,showpacs,floatfix,nofootinbib,notitlepage,twocolumn,longbibliography]{revtex4-1}
\usepackage{amsmath,graphicx,float,csquotes,mdframed,appendix,url}
\usepackage[dvipsnames]{xcolor}
\usepackage[colorlinks=true, pdfstartview=FitV, linkcolor=RedOrange, citecolor=Emerald, urlcolor=Cerulean]{hyperref}
\usepackage{amssymb}
\usepackage{soul}

\newcommand{\snn}{\sqrt{s_\mathrm{NN}}}
\newcommand{\glb}{\textsc{3d-glauber}}
\newcommand{\iebemusic}{\textsc{iebe-music}}
\newcommand{\music}{\textsc{music}}

\begin{document}

\preprint{}
\title{Unveiling baryon charge carriers through charge stopping in isobar collisions}

\author{Gregoire Pihan}
\email{GregoirePihan@wayne.edu}
\affiliation{Department of Physics and Astronomy, Wayne State University, Detroit, Michigan 48201, USA}

\author{Akihiko Monnai}
\affiliation{Department of General Education, Faculty of Engineering, Osaka Institute of Technology, Osaka 535-8585, Japan}

\author{Bj\"orn Schenke}
\affiliation{Physics Department, Brookhaven National Laboratory, Upton, NY 11973, USA}

\author{Chun Shen}
\affiliation{Department of Physics and Astronomy, Wayne State University, Detroit, Michigan 48201, USA}
\affiliation{RIKEN BNL Research Center, Brookhaven National Laboratory, Upton, NY 11973, USA}

\begin{abstract}
Utilizing a comprehensive (3+1)D relativistic hydrodynamic framework with multiple conserved charge currents and charge-dependent Lattice-QCD-based equation of state, we study the baryon and electric charge number deposition at mid rapidity in isobar Ru+Ru and Zr+Zr collisions at the center of mass energy $\sqrt{s_\mathrm{NN}}=200$ GeV. Comparing our predictions with upcoming experimental data from the Relativistic Heavy Ion Collider will shed light on the existence of baryon junctions.
\end{abstract}

\maketitle

\paragraph{1. Introduction.}

Relativistic nuclear collisions produce extreme conditions characterized by extraordinarily high temperatures and densities, providing an opportunity to probe the details of emergent many-body interactions within Quantum Chromodynamics (QCD)~\cite{Arslandok:2023utm, Achenbach:2024sch}.
The study of conserved charges, such as net-baryon number, in the environment created by such collisions allows us to elucidate properties of the strongly interacting quark-gluon system across the QCD phase diagram at finite densities~\cite{Bzdak:2019pkr, Monnai:2021kgu, An:2021wof}. 
Exploring the QCD phase diagram at large densities also establishes connections between heavy-ion physics and nuclear astrophysics, for example, related to recent neutron star merger observations \cite{Lovato:2022vgq, Sorensen:2023zkk}.

A crucial ingredient to understanding net-baryon transport in heavy-ion collisions is the stopping of the initial-state baryon charge. In addition to incoming nucleons losing their energy and momentum through collisions, their baryon charges are transported toward mid-rapidity. The baryon stopping can be enhanced by fragmentation mechanisms in QCD strings \cite{Andersson:1983ia, Vance:1997th, Sjostrand:2006za, Bahr:2008pv, Pratt:2023pee}, or in a picture where the baryon number is carried by the gauge-invariant string junctions inside a nucleon as opposed to the valence quarks \cite{Montanet:1980te, Kharzeev:1996sq}.

In heavy-ion collisions, the dynamical evolution of conserved charge currents, encompassing net electric charge, net baryon number, and net strangeness, is complex and coupled non-linearly through the QCD equation of state (EOS)~\cite{Monnai:2019hkn, Monnai:2021kgu, Karthein:2021nxe, Almaalol:2022xwv}. The relative chemical contents in the final-state hadronic measurements carry information on the conserved charges' space-time distributions at freeze-out~\cite{Denicol:2018wdp, Shen:2018pty, Du:2023gnv}. Particularly, the net baryon and electric charge rapidity distributions can elucidate how different conserved charges are transported along the longitudinal direction during the collision~\cite{Lewis:2022arg, Shen:2022oyg, Du:2022yok, Pihan:2023dsb}. This multiple conserved charge transport also allows for probing differences in the nuclear structure of the colliding  nuclei~\cite{Xu:2021qjw, Pihan:2023dsb}.

This work explores the (3+1)D dynamics of multiple conserved charge currents in relativistic collisions of isobaric nuclei. 
In high-energy collisions, it is crucial to understand how energy-momentum, net baryon number, and electric charges are distributed after the initial collision impact~\cite{Shen:2022oyg, Pihan:2023dsb}. 
By studying the difference between net baryon and net electric charge dynamics in relativistic isobar collisions at the Relativistic Heavy Ion Collider (RHIC)~\cite{Lewis:2022arg}, we explore the effects of having equal or different carriers for net baryon and net electric charge in nucleons. Comparison with the upcoming experimental data from the STAR Collaboration will reveal which situation is preferred and whether there are indications for a scenario where gluonic string junctions inside the nucleon carry the baryon charge~\cite{Montanet:1980te, Kharzeev:1996sq}.

We note that the theoretical framework developed in this work enables the exploration of the QCD phase diagram in four dimensions: temperature, net baryon, electric charge, and strangeness densities \cite{Monnai:2021kgu}. This opens up a wide range of studies, such as collisions with various $Z/A$ nuclei and center of mass energies, which we leave for future work. 

\paragraph{2. Theoretical framework.}

We simulate the (3+1)D space-time evolution of ruthenium+ruthenium (Ru+Ru) and zirconium+zirconium (Zr+Zr) collisions event-by-event using the \iebemusic{} framework~\cite{Shen:2022oyg}. The initial-state energy, momentum, and conserved charge transport are modeled by string deceleration within the \glb{}
model~\cite{Shen:2017bsr, Shen:2022oyg}. This work further extends the model to parameterize the system's net electric charge distribution after the initial impact. All the conserved quantities, namely the energy-momentum tensor, net baryon, and electric charges, are fed into hydrodynamic fields as source terms and evolved with \music{}~\cite{Schenke:2010nt, Schenke:2010rr, Paquet:2015lta, Denicol:2018wdp}. 

When the local energy density of fluid cells drops below $e_\mathrm{sw} = 0.35$~GeV/fm$^3$, the constant energy density surface elements are identified with the Cornelius algorithm \cite{Huovinen:2012is} and they are converted to hadrons using the Cooper-Frye particlization procedure~\cite{Cooper:1974mv, Shen:2014vra}. The produced hadrons are further evolved dynamically until kinetic freeze-out and decay to stable hadronic states through the hadronic transport model \textsc{urqmd}~\cite{Bass:1998ca, Bleicher:1999xi}. We include strong and weak decay contributions to the final stable hadrons. 

With the extended \glb{} model, we can model the independent initial-state stopping of baryon and electric charges. As the two colliding nuclei pass through each other,
strings form between quarks from two colliding nucleons and cause the quarks to decelerate. The lost energy and momentum are distributed along the string and become part of the produced medium. The baryon and electric charges from the participant nucleons are assigned to be carried by either the produced strings or the wounded nucleons' remnants with an equal probability. The detailed model implementation can be found in Ref.~\cite{Shen:2022oyg}.

When a baryon ($B$) or electric charge ($Q$) is assigned to a string, we assume that the charge's rapidity follows the probability distribution
\begin{equation}
    P(y^{X}) = (1 - \lambda_X) y_{P/T} + \lambda_X \frac{e^{[y^{X} - (y_P + y_T)/2]/2}}{4 \sinh{[(y_P - y_T)/4]}},
    \label{eq:GJprobability}
\end{equation}
where $X = B, Q$ and $y_{P/T}$ are the string ends' rapidities in the projectile- and target-going directions. The model parameter $\lambda_{X}$ controls the relative probability for the charge to 
follow the distribution obtained from the stopping of single string junctions \cite{Kharzeev:1996sq}.
Ref.~\cite{Shen:2022oyg} demonstrated for Au+Au collisions that Eq.~\eqref{eq:GJprobability} is an effective way to account for additional initial state baryon transport over the string deceleration.

In this work, we implement the same description for electric charges and sample their distribution with Eq.~\eqref{eq:GJprobability} independently from the incoming baryon charges. This implementation allows us to study final-state observables for the scenario ($\lambda_Q = \lambda_B$) that baryon and electric charges lose the same amount of energy during the initial impact of the two colliding nuclei. Regardless of the detailed implementation of the stopping mechanism, if experimental measurements prefer $\lambda_Q \neq \lambda_B$, it would suggest different initial-state stopping mechanisms for baryon and electric charges.

As we study collisions at top RHIC energy and observables at mid-rapidity, we do not need to perform the fully dynamical initialization \cite{Shen:2017bsr}, but feed all source terms into the hydrodynamic simulations at one given time $\tau = 0.5$~fm. 

Net baryon and electric charge currents are dynamically evolved along with energy and momentum by solving the conservation equations,
\begin{align}
    \partial_{\mu} T^{\mu \nu} &= 0\\
    \partial_{\mu} J_{X}^{\mu} &= 0,
    \label{Eq:idealHydroEoM}
\end{align}
where $T^{\mu \nu}$ is the energy-momentum tensor and ${J_{X}^{\mu} \equiv \rho_X u^{\mu}}$ denotes for conserved charge currents with $X = B,Q$. These hydrodynamic equations of motion are coupled through the equation of state, ${\mathcal{P} = \mathcal{P}(\epsilon, \rho_B, \rho_Q, \rho_S)}$. 
For this work, we developed a four-dimensional extension of the EOS model \textsc{neos} \cite{Monnai:2019hkn}, which is built from a Taylor expansion employing susceptibilities from Lattice QCD calculations up to the 4th order and matched to a hadron resonance gas model. In our simulations, we assume local strangeness neutrality, $\rho_S = 0$. 

During the hydrodynamic evolution, we consider shear and bulk viscous effects on the system's energy-momentum tensor. The shear and bulk viscous tensors are evolved according to the Denicol-Niemi-Molnar-Rischke (DNMR) hydrodynamic evolution equations \cite{Denicol:2012cn, Denicol:2018wdp}. The specific shear and bulk viscosities
are parametrized as
\begin{align}
    \frac{\eta T}{e + \mathcal{P}} &= \eta_0 \left[1 + s \left(\frac{\mu_B}{\mu_{B, 0}}\right)^a \right] \\
    \frac{\zeta T}{e + \mathcal{P}} &= \zeta_0 \exp \left[-\left(\frac{T - T_\mathrm{peak}}{T_{\mathrm{width}, \lessgtr}}\right)\right],
\end{align}
where ${T_{\mathrm{width},<} = 0.01}$~GeV for ${T < T_\mathrm{peak}}$, and ${T_{\mathrm{width},>} = 0.08}$~GeV for ${T > T_\mathrm{peak}}$ with ${T_\mathrm{peak} = 0.17}$~GeV
and $\zeta_0$ = 0.1. For the specific shear viscosity,
we set $\eta_0$ = 0.08, $s = 2$, $\mu_{B,0} = 0.6$~GeV, and $a = 0.7$.

We briefly return to the discussion of the modeling of the initial condition to emphasize that detailed nuclear structure information is crucial for understanding the observable ratios between the RHIC isobar collision systems~\cite{Xu:2017zcn, STAR:2021mii, Zhang:2021kxj, Nijs:2021kvn}. In particular, the electric charge-related observables are expected to be sensitive to the neutron skin of the colliding nuclei. To model the nuclear density, including the neutron skin, we consider spatial configurations of protons and neutrons with different Woods-Saxon profiles
\begin{equation}
    \rho_{p,n}(r, \theta, \phi) = 
    \frac{1}{1+e^{(r - R_{p,n}(\theta, \phi))/a_{p,n}}}
    \label{eq:WSprofile1}
\end{equation}
with $R_{p,n}(\theta, \phi) = R_{p,n}(1+\beta_2 Y_2^0(\theta,\phi) + \beta_3 Y_3^0(\theta,\phi))$. 
Here, the index $p,n$ denotes protons and neutrons, respectively, and $R_{p,n}$ is the half-density radius, $a_{p,n}$ is the surface diffuseness, and $\beta_{2,3}$ are the magnitudes of the quadrupole and octupole deformation parameters. The numerical values of all parameters used in this study are summarized in Table \ref{table:WSParams}.

\begin{table}[t!]
\centering
\caption{The Wood-Saxon parameters for ruthenium (Ru) and zirconium (Zr) used in this work~\cite{Xu:2021vpn, Zhang:2021kxj}.}
\begin{tabular}{|c | c | c | c | c | c | c |} 
 \hline
  & $R_p$ (fm)& $a_p$ (fm)&  $R_n$ (fm) & $a_n$ (fm) & $\beta_2$ & $\beta_3$ \\ 
  \hline
 Ru & 5.09 & 0.46 & 5.105 & 0.47 & 0.16 & 0 \\ 
 \hline
 Zr & 5.02 & 0.52 & 5.12 & 0.57 & 0.06 & 0.2 \\
\hline
\end{tabular}
\label{table:WSParams}
\end{table}

\paragraph{3. Results.}
We begin by constraining the model parameters related to initial-state baryon transport using net proton measurements in Au+Au collisions.
\begin{figure}[ht!]
    \includegraphics[scale=0.5]{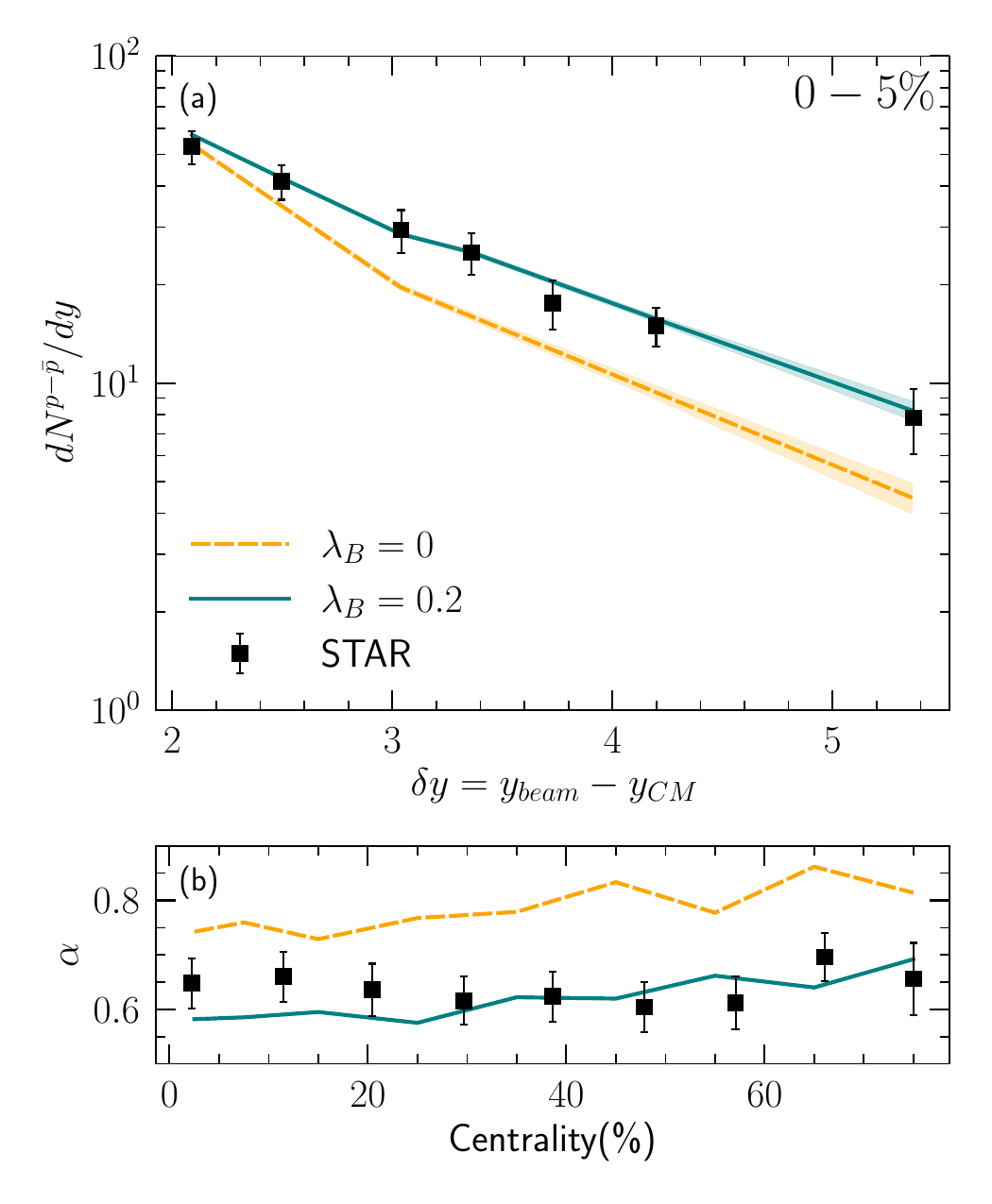}
    \caption{(Color online) \textit{Panel (a)}: The midrapidity net-proton yields for 0-5\% most central Au+Au collisions as a function of the rapidity loss $\delta y$. The STAR measurements~\cite{STAR:2008med, STAR:2017sal} are compared with theory results with $\lambda_B = 0$ and $\lambda_B = 0.2$.
    \textit{Panel (b)}: The fitted slope parameter $\alpha$ as a function of the centrality. (See text for the definition)}
    \label{fig:TuningLambdaB}
\end{figure}
Figure~\ref{fig:TuningLambdaB}(a) shows a comparison between the midrapidity net-proton yields measured at STAR~\cite{STAR:2008med, STAR:2017sal} (black squares) and the model results from the \iebemusic{} framework (solid and dashed lines) for 0-5\% most central Au+Au collisions from 7.7 to 200 GeV. The results are plotted as a function of rapidity loss $\delta y = y_\mathrm{beam} - y_\mathrm{CM}$, where the beam rapidity $y_\mathrm{beam} \equiv \mathrm{arccosh}(\snn/(2 m_N))$ and the center of mass rapidity $y_\mathrm{CM} = 0$ for Au+Au collisions. The nucleon mass $m_N = 0.938$~GeV and $\snn$ is the center of mass energy per nucleon pair in ${\rm GeV}$. The model results with $\lambda_B = 0.2$ in Eq.~\eqref{eq:GJprobability} show good agreement with the experimental data, while the results with $\lambda_B=0$ underestimate the measured net proton yields for $\snn \ge 10$~GeV. 

The collision energy dependence of the mid-rapidity net proton yields can be characterized by a fitted slope parameter $\alpha$ as in $dN^{p - \bar{p}}/dy \propto \exp(-\alpha \delta y)$. Figure~\ref{fig:TuningLambdaB}(b) shows $\alpha$ as a function of the collision centrality for theory calculations and the STAR measurements. Our simulations with $\lambda_B=0.2$ yield the slope parameter $\alpha \sim 0.6$, in line with the STAR measurements over all centrality bins. The fact that the STAR measurements disfavor the $\lambda_B = 0$ case in Eq.\,\eqref{eq:GJprobability} suggests that the baryon transport from string deceleration is not enough.
Additional charge stopping, as achieved for example by assigning the baryon charge to a gluonic string junction, is needed to reproduce the experimental measurements.

Because the Ru and Zr nuclei have the same mass number $A$ but different charge numbers $Z$, the RHIC isobar collisions allow one to study the baryon stopping dynamics relative to the electric charge transport with the following ratio~\cite{Lewis:2022arg}:
\begin{equation}
    r = \frac{(N_{B}^{\mathrm{Ru}} + N_{B}^{\mathrm{Zr}})}{2(N_{Q}^{\mathrm{Ru}} - N_{Q}^{\mathrm{Zr}})} \times \frac{\Delta Z}{A}\,.
    \label{eq:ratio}
\end{equation}
We compute the net baryon number of each collision system as $N_{B} \equiv (N_{p} - N_{\bar{p}}) + (N_{n} - N_{\bar{n}})$ and the net electric charge $N_{Q} \equiv (N_{\pi^+} - N_{\pi^-}) + (N_{K^+} - N_{K^-}) + (N_{p} - N_{\bar{p}})$. For the RHIC isobar collisions, we normalize the ratio of net baryon to electric charges by the factor ${\Delta Z/A = (Z^{\mathrm{Ru}} - Z^{\mathrm{Zr}})/A = 4/96}$. A ratio $r > 1$ indicates stronger stopping for baryon charges compared to electric charges.

\begin{figure}[t!]
    \includegraphics[scale=0.4]{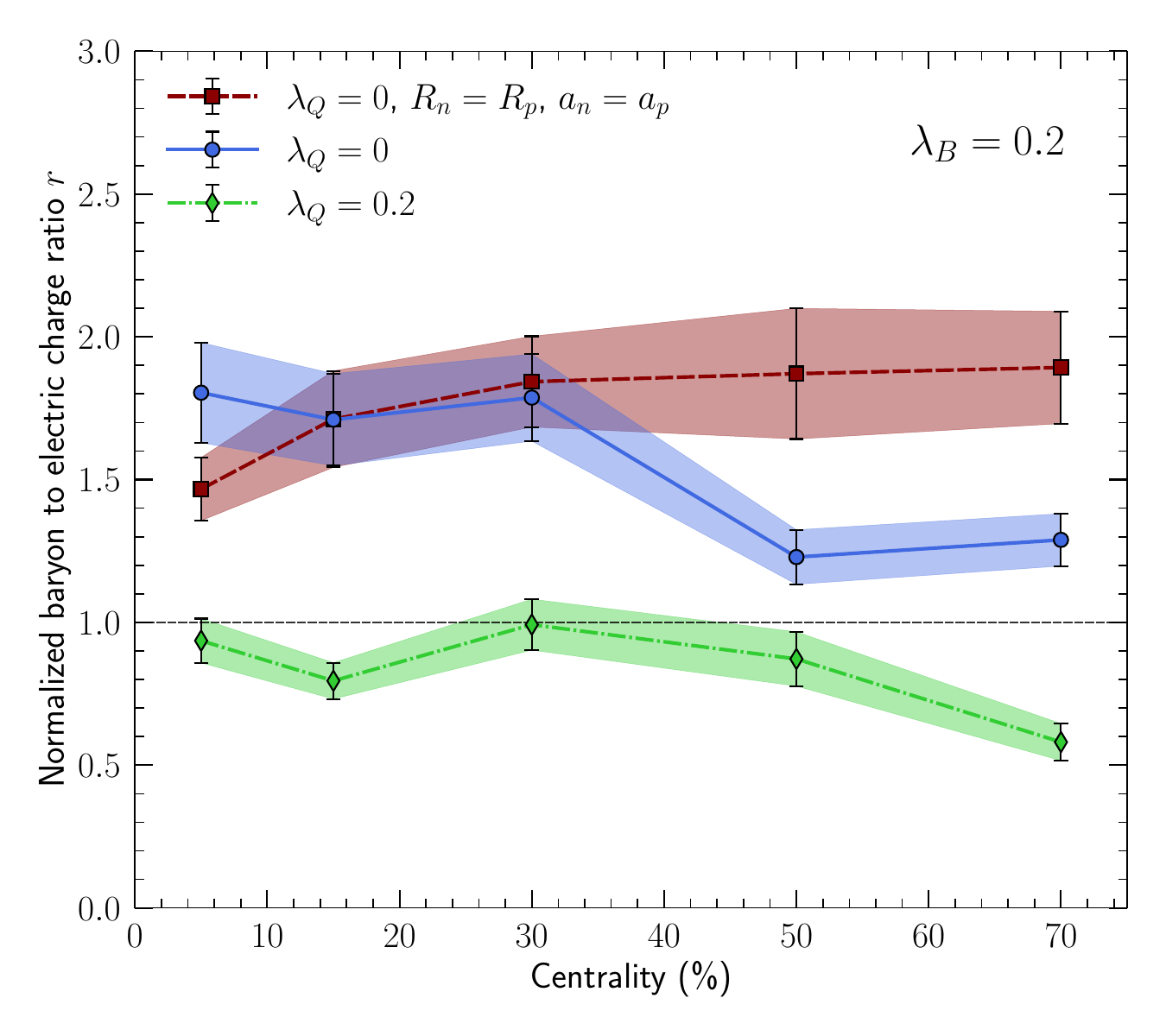}
    \caption{(Color online) The baryon to electric charge ratio $r$ for the RHIC isobar collisions as a function of centrality with different initial-state electric charge stopping.}
    \label{fig:FinalRatio}
\end{figure}

Figure~\ref{fig:FinalRatio} shows the normalized baryon to electric charge ratio $r$ (in Eq.~\eqref{eq:ratio}) as a function of the collision centrality defined by the charged hadron multiplicity in Ru+Ru collisions. We adjust the centrality boundaries of Zr+Zr collisions so that the two collision systems have the same averaged charged hadron multiplicity in every centrality bin. This re-binning of the Zr+Zr collision centrality is essential to eliminate the bias on the number of participants introduced by the nuclear structure difference in the two nuclear species. We verify that our simulations with equal baryon and electric charge stopping in the initial state ($\lambda_Q=\lambda_B=0.2$) reproduce a ratio $r \approx 1$ up to 50\% collision centrality. The ratio $r < 1$ in peripheral collisions beyond 50\% centrality can be understood as caused by the difference in neutron skin between the two different nuclei (see later discussion).

As we reduce the initial-state electric charge transport by requiring electric charges to remain at the string ends (in the valence quarks), which corresponds to $\lambda_Q = 0$ (solid line), the final-state baryon to electric charge ratio $r$ increases significantly. The value of $r$ stays roughly constant up to 30\% in centrality and decreases in more peripheral collisions. Because we constrained $\lambda_B$ in Au+Au collisions, this result presents a quantitative prediction whose experimental confirmation would strongly support the picture where baryon charges are carried by baryon junctions while electric charges rest in the valence quarks.

To understand the centrality dependence of the normalized baryon to electric charge ratio $r$, we perform simulations without neutron skin in the isobars (dashed line). In this case, the ratio $r$ stays constant in peripheral collisions. A finite neutron skin leads to more neutrons populating the edge of the nucleus. Because the events in peripheral centralities correlate to collisions with large impact parameters, there are more neutrons than protons among the participant nucleons, which reduces the total net electric charges available to be transported. Because the Zr nucleus has a larger neutron skin than the Ru nucleus, the net electric charge difference $\Delta N_Q = N_{Q}^{\mathrm{Ru}} - N_{Q}^{\mathrm{Zr}}$ is larger for peripheral collisions leading to a smaller ratio $r$. The same neutron skin effect on the centrality dependence of the baryon to electric charge ratio $r$ is present in the equal stopping case ($\lambda_Q = \lambda_B = 0.2$) discussed above.

\paragraph{4. Conclusion.}

Motivated by the possibility that the string junction in a nucleon could carry the baryon charge, we explore potential experimental observables in relativistic heavy-ion collisions sensitive to early-stage baryon transport. The collision energy dependence of the net proton yields in Au+Au collisions and the baryon to electric charge ratio in the isobar collisions can elucidate the early-stage baryon transport and its mechanism. 

We computed the net baryon number and net charge at midrapidity using the \iebemusic{} framework with the \glb{} initial state, an equation of state that depends on temperature and three conserved charge densities (net baryon, electric charge, and strangeness), and final state hadronic rescattering. We employed a baryon-stopping probability inspired by the rapidity dependence of baryon junction stopping and constrained by the measured net-proton distribution in Au+Au collisions. Comparing two scenarios, one using the same description for electric charge stopping, the other corresponding to electric charge following the valence quarks, we found a dramatic difference in the scaled ratio of net baryon number to electric charge difference between the isobars at midrapidity. Agreement of experimental data with the prediction for different stopping mechanisms will strongly support a scenario in which the baryon number is not carried by the valence quarks.

Another striking result of our analysis is the studied observable's strong sensitivity to the colliding nuclei's neutron skin. This demonstrates that measurements of electric charge transport in isobar collisions can be used to constrain the neutron skin very directly, without the need for measuring, e.g.~parity-violating asymmetries in polarized electron scattering~\cite{PREX:2021umo}, or extracting the neutron skin from a constrained nuclear mass distribution ~\cite{Giacalone:2023cet}.

\paragraph{Acknowledgments}
We thank Nicole Lewis, Scott Pratt, Prithwish Tribedy, and Zhangbu Xu for useful discussions.
This work is supported by the U.S. Department of Energy, Office of Science, Office of Nuclear Physics, under DOE Contract No.~DE-SC0012704 and within the framework of the Saturated Glue (SURGE) Topical Theory Collaboration (B.P.S.) and Award No.~DE-SC0021969 (C.S. \& G.P.). This work is also supported by JSPS KAKENHI Grant Number JP24K07030 (A.M.).
C.S. acknowledges a DOE Office of Science Early Career Award. 
This research was done using computational resources provided by the Open Science Grid (OSG)~\cite{Pordes:2007zzb, Sfiligoi:2009cct}, which is supported by the National Science Foundation award \#2030508.

\bibliography{biblio}

\end{document}